\documentclass[a4paper]{llncs}

\usepackage{graphicx}
\usepackage{amsmath}
\usepackage[T1]{fontenc}
\usepackage{hyperref}
\usepackage{varioref}
\usepackage{xspace}
\usepackage{paralist}
\usepackage{fancybox}
\usepackage{xcolor}
\usepackage{calc}
\usepackage{verbatim}

\usepackage{relsize}
\usepackage{booktabs}
\usepackage[scaled]{helvet}

\usepackage{times}

\usepackage{listingsVDM}
\usepackage{overturelanguagedef}
\lstdefinestyle{overtureLanguageStyle}{basicstyle=\ttfamily,
			frame=trBL,
			showstringspaces=false,
			frameround=fttt,
			captionpos=b,
			aboveskip=5mm,
			belowskip=5mm,
			framexleftmargin=0mm,
			framexrightmargin=0mm}

\usepackage{customlangdef}
\definecolor{maroon}{rgb}{0.5,0,0}
\definecolor{darkgreen}{rgb}{0,0.5,0}
\definecolor{ao}{rgb}{0.0, 0.5, 0.0}
\definecolor{mycolor1}{rgb}{0.0, 0.53, 0.74}%
\definecolor{mycolor2}{rgb}{0.21783,0.72504,0.61926}%
\definecolor{mycolor4}{rgb}{0.93, 0.53, 0.18}%
\definecolor{plum}{rgb}{0.56, 0.27, 0.52}
\definecolor{pinegreen}{rgb}{0.0, 0.47, 0.44}
\definecolor{pthaloblue}{rgb}{0.0, 0.06, 0.54}
\definecolor{saffron}{rgb}{0.96, 0.77, 0.19}

\usepackage[caption=false]{subfig} 
\usepackage{pgfplots}
\pgfplotsset{compat=newest}
\usetikzlibrary{pgfplots.groupplots, pgfplots.units}
\pgfplotsset{
	every axis label/.append style={font=\normalsize},
	tick label style={font=\small},
	/pgfplots/enlargelimits=false,
    legend style={legend pos=north east, font=\small},
    legend cell align=left,
    xlabel near ticks,
    ylabel near ticks,
	axis on top,
    highlight/.code args={#1:#2}{
        \fill [every highlight] ({axis cs:#1,0}|-{rel axis cs:0,0}) rectangle ({axis cs:#2,0}|-{rel axis cs:0,1});
    },
    /tikz/every highlight/.style={
        on layer=\pgfkeysvalueof{/pgfplots/highlight layer},
        red!10
    },
    /tikz/highlight style/.style={
        /tikz/every highlight/.append style=#1
    },
    highlight layer/.initial=axis background
}%

\lstdefinelanguage{XML}
{
  basicstyle=\ttfamily\scriptsize,
  morestring=[b]",
  moredelim=[s][\bfseries\color{maroon}]{<}{\ },
  moredelim=[s][\bfseries\color{maroon}]{</}{>},
  moredelim=[l][\bfseries\color{maroon}]{/>},
  moredelim=[l][\bfseries\color{maroon}]{>},
  morecomment=[s]{<?}{?>},
  morecomment=[s]{<!--}{-->},
  commentstyle=\color{darkgreen},
  stringstyle=\color{blue},
  identifierstyle=\color{red}
}

\usetikzlibrary{shapes,fit,arrows,calc,arrows,arrows.meta,positioning}

\usepackage{morten_preamble}

\begin{document}

\title{Modelling Chess in VDM++}
\titlerunning{Modelling Chess in VDM++}

\author{Morten Haahr Kristensen\inst{1} and Peter Gorm Larsen\inst{1}  }
\institute{DIGIT, Department of Electrical and Computer Engineering, Aarhus University, Denmark, \email{201807664@post.au.dk, pgl@ece.au.dk}
}

\maketitle

\ifnotes %
\textbf{NOTES ENABLED. TURN OFF IN FINAL VERSION.} %
\else %
\fi %

\begin{abstract}
The game of chess is well-known and widely played all over the world. However, the rules for playing it are rather complex since there are different types of pieces and the ways they are allowed to move depend upon the type of the piece. In this paper we discuss alternative paradigms that can be used for modelling the rule of the chess game using VDM++ and show what we believe is the best model. It is also illustrated how this model can be connected to a standard textual notation for the moves in a chess game. This can be used to combine the formal model to a more convenient interface.
\end{abstract}

\section{Introduction}
\label{sec:intro}


Many games that humans can play with each other include rules based on logic about what is allowed. Board games often have different kinds of pieces that the players take turn in moving. One of the more complex games is called Chess. In this paper we model the game of chess using VDM++ and discuss the pros and cons of alternative modelling styles \cite{Fitzgerald&05}. The model was written as an educational example and can be executed to validate whether the rules of a chess game were broken.

The purpose of formal models is to enable formal analysis of desirable properties of the system of interest. However, with each formal model, there are abstraction alternatives and for each of these, it is worthwhile discussing the best paradigms for describing the system in question. In a specification focus is on explaining the key aspects in relation to the purpose of the model and in this context explainability to human beings is much more important than the speed of execution. The explainability of different paradigms will be discussed in this paper using the rules of the game of Chess as an example. 

This paper is structured as follows: After this introduction Section~\ref{sec:background} provides the reader with a brief introduction to the rules of the Chess game. Afterwards, Section~\ref{sec:paradigms} presents reflections about modelling chess using either the object-oriented or the declarative paradigms. Then Section~\ref{sec:coremodel} provides the core of the VDM++ model using the declarative paradigm. After that Section~\ref{sec:PGN} briefly explains how the Portable Game Notation can be incorporated enabling one to incorporate a standard textual format as input for the VDM++ model making it easy to take existing games that have been played into the model. Section~\ref{sec:related} briefly relates the contribution of this paper with related work.
Finally, Section~\ref{sec:concl} provides a few concluding remarks and considers the future directions.

\section{The Rules of Playing Chess}
\label{sec:background}

%

Chess is a two-player turn-based game where one player controls the white pieces and the other controls the black pieces. The game is played on a chessboard that consists of 64 squares (fields) shaped in an 8x8 grid. The columns of the grid are called ``files'' and the rows are called ``ranks''. The squares are coloured alternatively light and dark and a line of squares of the same colour going from one edge to an adjacent edge on the board is called a ``diagonal''. The squares following the horizontal axis on the board are annotated using letters the 'a' -- 'h' and the vertical squares are annotated using the numbers 1 -- 8.

\begin{figure}[hbt]
  \centering
  \includegraphics[width=0.5\textwidth]{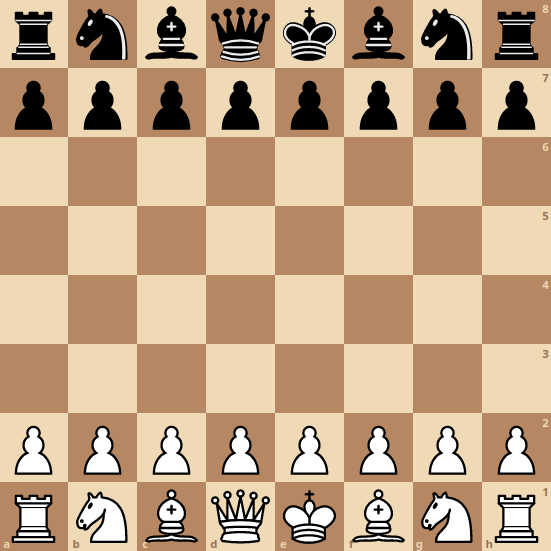}
  \caption{Initial position of the chess pieces where the first row with squares 'a' -- 'e' shows the position of respectively a rook, knight, bishop, queen and king. The second row shows eight white pawns. Image source: \cite{ChessBoardEditor}.}
  \label{fig:coremodel:board_initial}
\end{figure}

Each player initially controls 16 pieces on the board, as seen on \cref{fig:coremodel:board_initial}, where a piece can be discriminated through three different attributes. The first is the piece type, which determines how a piece is allowed to move. The second is the position which determines which square a piece is placed on. Finally, each piece has a colour indicating which player controls it.

There are six types of pieces in the game where each player initially controls eight pawns, two rooks, two knights, two bishops, a queen, and, most importantly, a king.
Each piece type has a unique way of moving around the board. For example the rook can move along either the file or rank on which it stands and the bishop can only move along the diagonals.
A piece can be blocked from moving to a square if there is another piece between the initial and the desired square. However, if an enemy piece occupies the desired square, the player may capture that piece by removing it from the board. When a player can capture a piece, the piece is said to be under attack. The knights have an exception in the moving pattern since they are not blocked by other pieces in their path.

The goal of the game is to put the king of the opposing player in a position where it is impossible to prevent it from being captured in the following turn\footnote{For a full list of rules see
\url{https://www.fide.com/FIDE/handbook/LawsOfChess.pdf}.} 

There are certain types of moves in chess that can be considered ``special'' in the regard, that they can only be performed when certain conditions are met. A simple example of such is that a pawn has the option of moving two squares forward from its initial position. Another example is ``castling'' which is the only type of movement involving two pieces. Castling allows the player to move their king two squares towards a rook on the player's first rank, then move the rook to the square the king just passed. However, castling is only allowed if the king and the rook have not been moved in during the game, if the squares between the king and rook are unoccupied and not under attack, and if the king is not under attack (in check).

If a pawn reaches the rank furthest from its initial position it must change its type to one of the
following: Queen, knight, rook or bishop. This type of move is called a ``promotion''. Taking this into account is actually challenging because it means that the type of that piece is changing dynamically.

\section{Alternative Paradigms for Modelling the Game of Chess}
\label{sec:paradigms}

The different VDM dialects generally encourage following a Functional Paradigm (FP), but the VDM++ dialect introduces the possibility of using Object-Oriented Paradigm (OOP) for structuring the models. In an OOP setting, there is typically a need to have instance variables inside classes to represent state, and to access and adjust these there is typically a need for operations that need to use the imperative paradigm with assignments to such instance variables. The question of whether or not to use such features arise.

When determining a paradigm to follow, one must consider if it is easier to encapsulate the moving parts or to minimise them. In the case of the game of Chess both OOP and FP paradigms may look appealing as many of the rules of chess generally are stateless and therefore without moving parts. However, the special rules in particular introduce statefulness to the game. This includes example moves such as en passant and castling, where the validity of the moves depends on previous moves.

\subsection{Considering the Functional and Object-Oriented Paradigms}

\begin{figure}[hbt]
    \centering
    \includegraphics[width=0.5\textwidth]{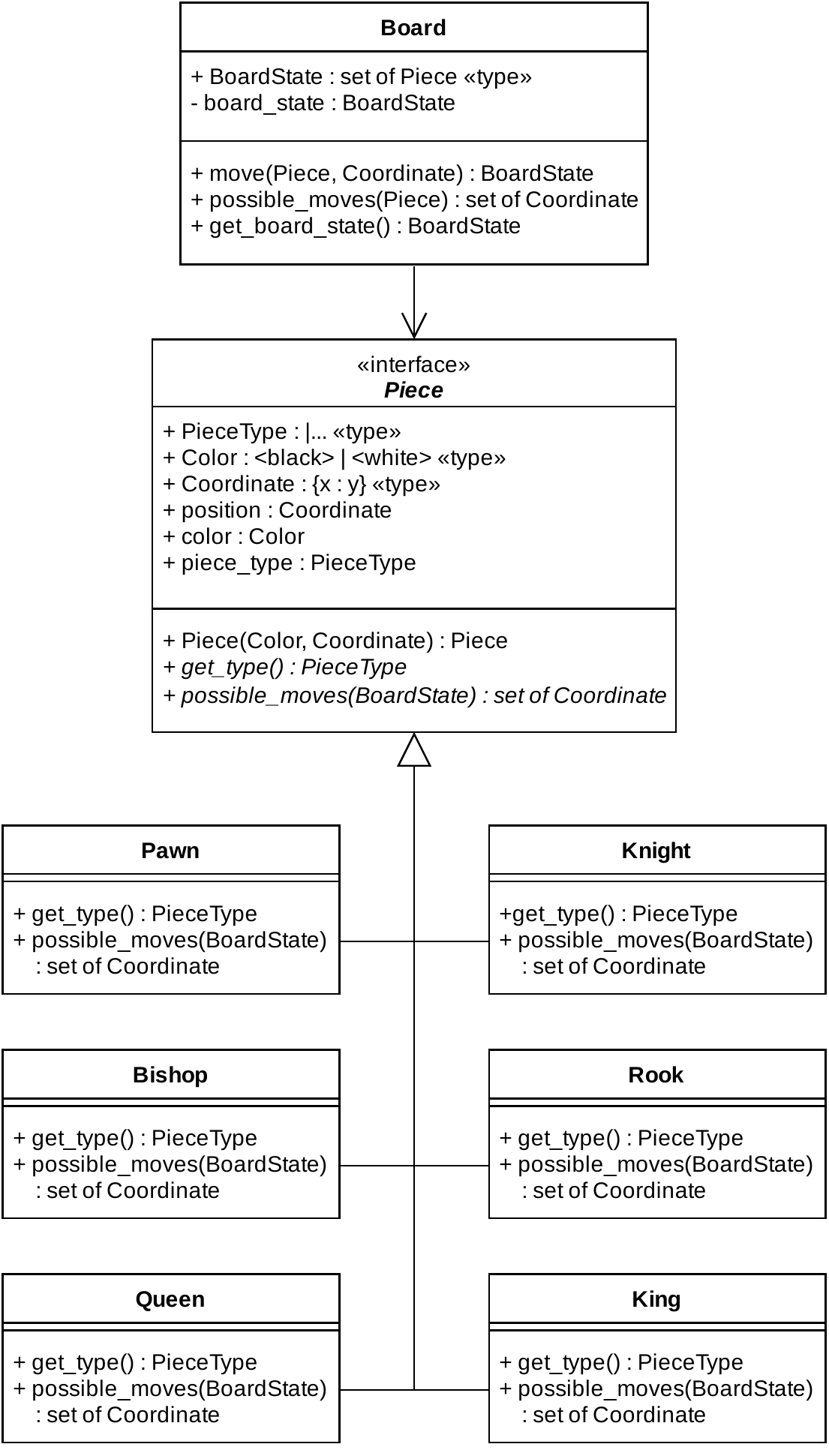}
    \caption{UML class diagram of the basis for an object-oriented architecture for chess.}
    \label{fig:par:op_arch}
\end{figure}

Two architectures following OOP and FP were considered. The first follows a typical OOP structure with a base class \texttt{Piece} that defines basic methods for determining possible moves. Each piece type then has a subclass implementation defining its unique movement pattern and potential attributes. The special moves would be modelled through \texttt{Board\`{}possible\_moves} as the \texttt{Board} class knows the state of the game. A simplified class diagram of such an architecture can be seen in \cref{fig:par:op_arch}.

\begin{figure}[hbt]
    \centering
    \includegraphics[width=0.5\textwidth]{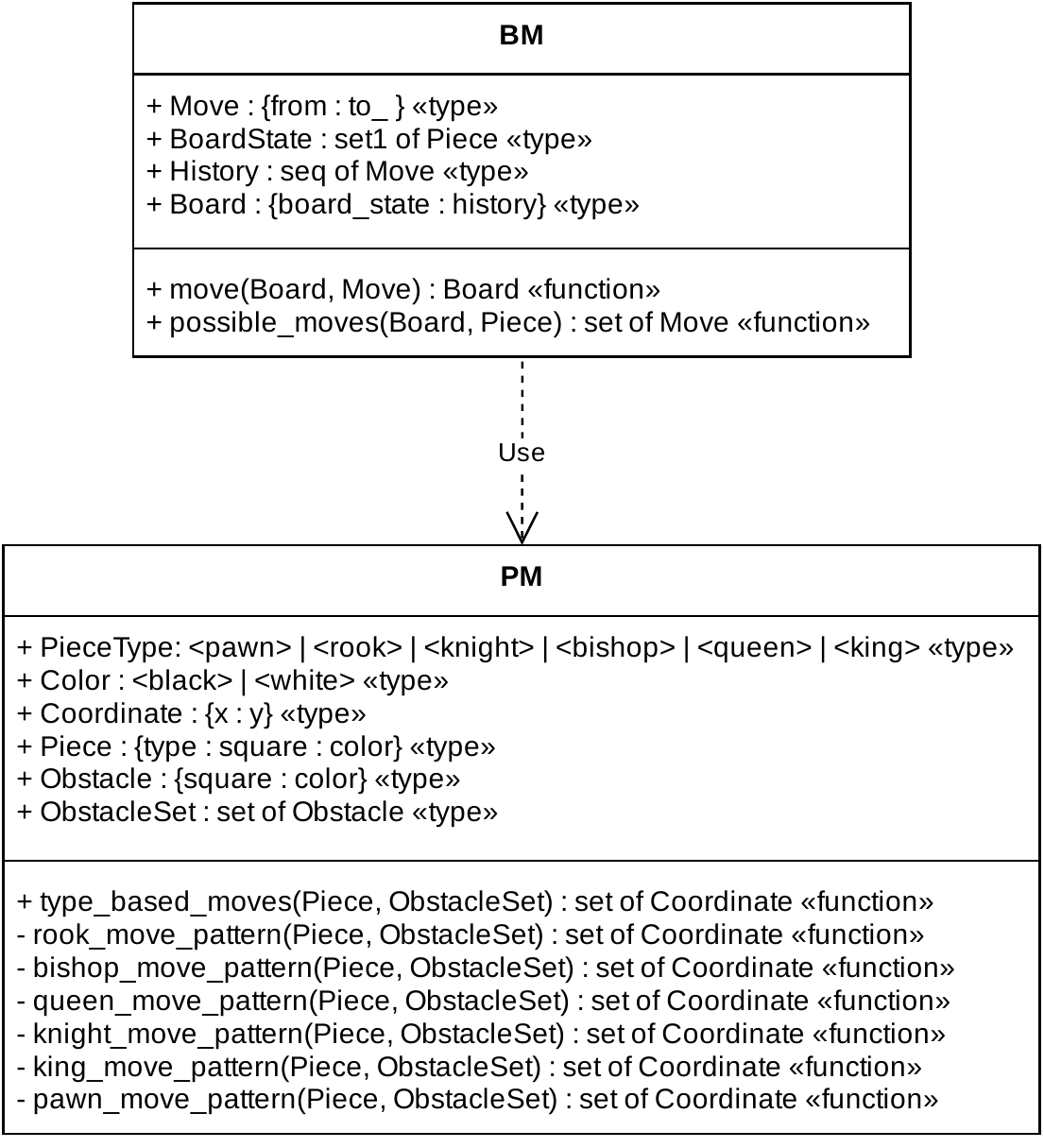}
    \caption{UML class diagram of the basis for a functional-style architecture for Chess.}
    \label{fig:par:fp_arch}
\end{figure}

The second architecture was written following the FP where only immutable variables were used. In VDM++ this meant defining all data structures as composite types. The benefit of this is that one only has values in the specification which means that the model is stateless. When writing VDM++ in a functional style, one should consider the classes as modules that encapsulate functionalities together in a namespace (it would thus look more like a VDM-SL model but we have kept it as a VDM++ model in order to ease the comparison). An example of such an architecture can be seen in \cref{fig:par:fp_arch}. In this architecture, all pieces share a common data structure containing their colour, position and type. 

One benefit of following the FP architecture relates to reasoning about the model. Generally speaking, it can be more difficult to reason about imperative models as their functionality may depend on a global model state. In order to formally verify the behaviour of a function inside an imperative model, one must consider all the methods that can also manipulate the global state. In contrast, a purely functional model consists of referentially transparent functions. 

Another benefit relates to testing and verification of the model. When states are introduced to a model the complexity increases significantly as there are more moving parts and thereby more test combinations to be written if the model is to be tested exhaustively. In practice, this means that a model following the FP requires fewer tests to be written as one does not need to consider all the states the model may appear in.

However, modelling Chess through an FP architecture also has some limitations, in particular related to the stateful aspects of the game e.g., when modelling castling one needs knowledge of whether the involved king and rook have moved. To determine this with FP one must know the previous moves made in the game and determine if the pieces were involved. With OOP one can simply introduce a boolean attribute on the rooks and kings indicating whether they have moved. While this is also possible following the FP, it implicates all the pieces as inheritance is not available.

At this point, it should be clear that there are benefits to both types of architecture. One should consider whether it is more important to minimise or encapsulate the moving parts when writing the architecture, as this decision may greatly impact the difficulty of the implementation. Finally, one must also consider how exhaustively the model is to be tested as a stateless model should contain less testing combinatorics compared to a stateful model.

An example of how FP can be more elegant than OOP can be found when moving pieces. It is to be assumed that a \texttt{Move} is modelled as a composite type of two \texttt{Piece}s, one indicating which \texttt{Piece} is being moved and the other indicating where it is moved to\footnote{The latter must be a \texttt{Piece} over a \texttt{Coordinate} to account for promotion.}, and that the \texttt{BoardState} is a \texttt{set of Piece}s.
In the OOP case, one would need to find the \texttt{Piece} in the current \texttt{BoardState} with similar attributes as \texttt{Move.from\_}, check if a \texttt{Piece} is occupying the same square as \texttt{Move.to\_}, potentially remove that \texttt{Piece}, and update the found \texttt{Piece} to match the new \texttt{Coordinate}. In the case of promotion, one would need to add a new \texttt{Piece} of the promoted type and remove the original.
When following FP one can simply update the \texttt{BoardState} by filtering out the \texttt{Piece}s with the \texttt{Coordinate}s in the \texttt{Move} and making a union with \texttt{Move.to\_}.

\subsection{Invariants on Compound Types in VDM++}

During the early stages of development where the OOP architecture was used, an issue within the VDM tool ``VDMJ'' \cite{Battle09} lead to an interesting discussion, that at its core relates to the choice of paradigm. The issue relates to the model snippet seen below where a \texttt{Piece} is moved on to a given \texttt{Coordinate}. The operation updates the state of the \texttt{Board} by first removing the captured \texttt{Piece} and then updating the position of the moved \texttt{Piece}. Since class instances are reference types in VDM++ the latter can be done through the assignment operator directly on the \texttt{Piece} as the \texttt{board\_state} has a reference to the instance. The \texttt{board\_state} is a \texttt{set of Piece} with an invariant stating that the positions of the \texttt{Piece}s inside the set must be unique.

\begin{lstlisting}
public move: Piece * Piece`Coordinate ==> ()
move(piece, coord) == (
    let current_coords = state_to_coords_set(board_state) in
        if coord in set current_coords then
            let dead_piece = {p | p in set board_state & p.position = coord} in
                board_state := board_state \ dead_piece;
    piece.position := coord
)
pre piece in set board_state and dead_piece in set board_state;
\end{lstlisting}

The logic of the model seemed sound but in practice, the interpreter would report an invariant violation when a \texttt{Piece} was captured. When debugging the operation it was shown that the invariant was correctly checked when updating the position of the \texttt{Piece}, but it was checked against an old \texttt{board\_state} still containing the \texttt{dead\_piece}, which caused the invariant to be violated.
In short, the issue was caused by an ``invariant listener'' on the \texttt{piece} object that was not correctly updated when the \texttt{board\_state} was modified\footnote{A link to the issue can be found here:\\\url{https://github.com/overturetool/vdm-vscode/issues/197}}. While the exact issue is not a concern of this paper, the complexity of having invariants on compound types containing references is an interesting topic that showcases some of the issues mutability brings. 

Since VDM++ objects are references they bring aliasing to the language, i.e., an object like \texttt{piece} can be accessed and modified in several places in the specification. If such an object reference is also a member of a compound type instance with an invariant, e.g., \texttt{board\_state}, the invariant for the compound type must be checked whenever the object is modified. The invariant must also be checked when the compound type instance itself is modified.
Furthermore, an object like \texttt{piece} could potentially be placed in multiple instances of different compound types with different invariants, which would mean modifying \texttt{piece} would result in several invariants being checked. In the case of Chess this might not be an issue as the data structures are relatively flat but in complex industrial cases it may not be the case. The reason for this issue showing up is that in order to make the VDM interpreter efficient while still ensuring that invariants are not violated it only tests the invariants whenever changes are made where the invariants are used. In order for this to work in the presence of aliasing this essentially requires checking the transitive closure of instances connected. This is obviously not efficient so the VDM interpreter ignores such invariants going across instances and thus this is a challenge for the OOP model presented here. 

Had the specification been written using immutable datatypes like composite types, the invariant issue would not have been a concern, as the \texttt{BoardState} would then be a set containing immutable values instead of references. This means that the invariant would only need to be checked when the \texttt{board\_state} was changed (or following an entirely FP, when a new \texttt{BoardState} was created).

While the writer of a specification should not typically concern themselves with the details of the tools, we believe this example makes a strong argument of why following the OOP may introduce unnecessary complexity to a specification. If one was to reason about a specification following the OOP with an instance of a compound type, one would need to consider all the places where the instance was modified but also all the places where the members of the instance could be modified. All of this is without considering more complex OOP concepts such as inheritance that only strengthens the point.

\section{Overview of the FP VDM Model of Chess}
\label{sec:coremodel}


The model described in this section follows the FP architecture as shown in \cref{sec:paradigms}, where a bottom-up approach will be taken of the most interesting parts of the model.

\subsection{PieceModule}
The PieceModule (\texttt{PM}) class has the responsibility of defining the types necessary to describe a Chess piece and providing the functions necessary to describe their basic movement patterns.

The piece type can be modelled as a union of quote types with the following options: \texttt{<pawn>|<rook>|<knight>|<bishop>|<queen>|<king>}. A composite type is used to define the \texttt{Coordinate}s which consists of two \texttt{nat1} to describe the x- and y-coordinates of a piece. The convention of annotating the x-axis through letters is abstracted away to make the two axes consistent. An invariant is put on \texttt{Coordinate} to ensure that the position is legal, i.e., the values are less than nine.

A data structure for the \texttt{Piece}s could now be established as a composite type containing the attributes \texttt{type : PieceType}, \texttt{square : Coordinate}, and \texttt{colour : Colour}.

The simple moves, i.e., excluding special moves, can be found for a given \texttt{Piece} using the \texttt{type\_based\_moves} function that takes \texttt{Piece * ObstacleSet} as parameters and returns a \texttt{set of Coordinates} containing the valid \texttt{Coordinate}s that the \texttt{Piece} can move to.
An \texttt{ObstacleSet} is needed to filter out the moves that are invalid due to the \texttt{Piece}'s path being blocked by an \texttt{Obstacle}. An \texttt{Obstacle} contains a \texttt{Coordinate} and \texttt{Colour} where the \texttt{Colour} is needed to indicate if the \texttt{Obstacle} is capturable or not.
Alternatively, one could have modelled the function without the \texttt{ObstacleSet} by returning the collection of legal \texttt{Coordinate}s assuming the board was empty since this would decouple the \texttt{PieceModule} further from the state of the board. However, it would require the \texttt{BoardModule} to be more closely coupled to the \texttt{PieceType} as it would need a special case for handling the movements of a pawn, as the pawn moves forward but attacks diagonally.

The \texttt{type\_based\_moves} function is written using a cases expression to pattern-match the \texttt{PieceType} to a function defining the specific movement pattern of the piece. The movement pattern of the knight and king can be modelled using a common helper function, \texttt{possible\_move\_direction}, as seen in the snippet below. Among the \texttt{Piece} and \texttt{ObstacleSet}, a pair of ints are provided as parameters to indicate the direction where the possible moves are to be considered. A new \texttt{[Coordinate]} is generated based on the \texttt{dir} through \texttt{coordinate\_factory} that returns \texttt{nil} if the provided inputs result in an invalid \texttt{Coordinate}. \texttt{possible\_move\_direction} evaluates to \texttt{nil} if the generated \texttt{Coordinate} is invalid or occupied by a friendly piece. Otherwise, the new \texttt{Coordinate} is returned.

\begin{lstlisting}
possible_move_direction: Piece * ObstacleSet * (int * int) -> [Coordinate]
possible_move_direction(p, os, dir) ==
let new_c = coordinate_factory(p.square.x + dir.#1, p.square.y + dir.#2) 
in
  if (new_c = nil) or 
     exists piece in set os &
      (piece.square = new_c) and (piece.colour = p.colour) 
  then nil
  else new_c;
\end{lstlisting}

Having now defined \texttt{possible\_move\_direction} the movement pattern of the knight can be defined as below.

\begin{lstlisting}
knight_move_pattern : Piece * ObstacleSet -> set of Coordinate
knight_move_pattern(p, os) ==
  {possible_move_direction(p, os, mk_(1, 2)), -- 2Up1Right
   possible_move_direction(p, os, mk_(-1, 2)), -- 2Up1Left
   possible_move_direction(p, os, mk_(1, -2)), -- 2Down1Right
   possible_move_direction(p, os, mk_(-1, -2)), -- 2Down1Left
   possible_move_direction(p, os, mk_(2, 1)), -- 1Up2Right
   possible_move_direction(p, os, mk_(-2, 1)), -- 1Up2Left
   possible_move_direction(p, os, mk_(2, -1)), -- 1Down2Right
   possible_move_direction(p, os, mk_(-2, -1)) -- 1Down2Left
    } \ {nil} 
pre p.type = <knight>;
\end{lstlisting}

Similarly, the other piece types rook, bishop and queen has been modelled using the \texttt{possible\_moves\_direction} as seen in the snippet below. This time recursion is needed as these can continue in a direction they are blocked or can capture a piece.
Once again a new \texttt{[Coordinate]} is generated.
The first conditional is identical to the one in \texttt{possible\_move\_direction} but the empty set is returned in this case. A check is then made to determine if an opponent is on the square that is being evaluated, where the recursion is terminated and a set containing the \texttt{Coordinate} is returned. At last, the recursive case is defined where a union between the set containing the \texttt{Coordinate} and the result of recursively calling the function in the same direction is returned.

\begin{lstlisting}
possible_moves_direction: Piece * ObstacleSet * (int * int) -> set of Coordinate
possible_moves_direction(p, os, dir) ==
  let new_c = coordinate_factory(p.square.x + dir.#1, p.square.y + dir.#2) 
  in
     if (new_c = nil) or 
        exists piece in set os &
                (piece.square = new_c) and (piece.colour = p.colour) 
    then {}
    elseif exists piece in set os &
                (piece.square = new_c) and (piece.colour = opposite_color(p.colour))
     then {new_c}
     else {new_c} union 
            possible_moves_direction(mk_Piece(p.type, new_c, p.colour), os, dir);
\end{lstlisting}

\noindent In principle this recursive function should have a proper \texttt{measure} ensuring the termination but it is not straightforward.

The movement pattern of the queen can then be defined as seen below.

\begin{lstlisting}
queen_move_pattern : Piece * ObstacleSet -> set of Coordinate
queen_move_pattern(p, os) ==
  dunion {possible_moves_direction(p, os, mk_(0, 1)),
          possible_moves_direction(p, os, mk_(0, -1)),
          possible_moves_direction(p, os, mk_(1, 0)),
          possible_moves_direction(p, os, mk_(-1, 0)),
          possible_moves_direction(p, os, mk_(1, 1)),
          possible_moves_direction(p, os, mk_(-1, -1)),
          possible_moves_direction(p, os, mk_(-1, 1)),
          possible_moves_direction(p, os, mk_(1, -1))}
pre p.type = <queen>;
\end{lstlisting}

\subsection{BoardModule}
The BoardModule (\texttt{BM}) class has the responsibility of defining and updating the state of a chessboard. Furthermore, it determines whether or not the special rule moves are possible.

The first type defined in the BM class is the composite type \texttt{Move}. Initially, \texttt{Move} was modelled as a product type consisting of a \texttt{Piece * Coordinate}\footnote{\texttt{Move} was changed from a product type to a composite type as the intent is clearer when the fields are named compared to referencing them through ''.\#1'' and ''.\#2''.}. This structure made sense up until the point where promotion was implemented since promotion allows for the \texttt{PieceType} to be changed, which could not be captured with the old definition.
Instead, \texttt{Move} was modelled with the attributes \texttt{from\_} and \texttt{to\_} that are both of type \texttt{Piece}. An invariant was placed upon \texttt{Move} that states the following: \texttt{m.from\_.colour = m.to\_.colour and m.from\_.square <> m.to\_.square} \\since a \texttt{Move} cannot change the colour and must update the position of the \texttt{Piece}.

It was then possible to define \texttt{History} as a sequence of \texttt{Move}s. A sequence was chosen since the ordering of the moves matters and there might be duplicates if a player moves a piece back and forth.

A \texttt{BoardState} could then be defined as a \texttt{set1 of Piece}, where a set was chosen since the ordering does not matter and duplicates are not allowed as that would indicate two pieces of the same colour and type being placed on the same square. To restrict two \texttt{Piece}s from sharing a position the following invariant was written: \texttt{forall p1, p2 in set b \& p1 <> p2 => p1.square <> p2.square}.
Finally, a \texttt{Board} type was introduced as a composite type containing a \texttt{BoardState} and a \texttt{History}.

The function \texttt{possible\_moves} is responsible for finding the set of valid moves for a \texttt{Piece} where both the simple- and special movement patterns are considered. The function finds the set of simple type-based moves, the set of stateful special moves, and the set of illegal stateful moves. It then evaluates to the union between the former two with the set difference of the latter. The logic can essentially be boiled down to ``find the entire set of possible moves and remove the impossible ones''. Here a set comprehension is used to convert the \texttt{Coordinate}s from \texttt{simple\_moves} to \texttt{Move}s through the helper function \texttt{piece\_coord\_to\_move}.

\begin{lstlisting}
public possible_moves : Board * PM`Piece -> set of Move
possible_moves (board, piece) == (
  let state_p_moves = stateful_possible_moves(board, piece),
      state_imp_moves = stateful_impossible_moves(board, piece),
      simple_moves = PM`type_based_moves(piece,
          PM`pieces_to_obstacles(board.board_state)) in
          ({piece_coord_to_move(piece, c) | c in set simple_moves} union
              state_p_moves) \ state_imp_moves
)
pre piece in set board.board_state;
\end{lstlisting}

The function \texttt{stateful\_impossible\_moves} yields the \texttt{set of Move} containing \texttt{Move}s that result in the player losing, as the rules of Chess disallow such a move from being performed. Thus it contains the moves that put the player in check and if the player already is in check it filters out moves that do not put them out of check.
Furthermore, if the \texttt{PieceType} is \texttt{<pawn>} and promotion is possible then it also contains the \texttt{Move} where the pawn moves to a square on the last rank without changing the \texttt{PieceType}, as it is illegal to not promote the \texttt{Piece}.

\texttt{stateful\_possible\_moves} is a dispatcher that considers the special rules of the pieces, i.e., castling, promotion, en passant and the option for a pawn to move two squares on its first move.

\begin{lstlisting}
stateful_possible_moves: Board * PM`Piece -> set of Move
stateful_possible_moves(board, piece) == (
  cases piece.type:
      <pawn> -> dunion {
          pawn_move_two(board.board_state, piece),
          en_passant(board, piece),
          pawn_promotion(board.board_state, piece)},
      <king> -> castling_possible(board, piece),
      others -> {}
  end
);
\end{lstlisting}

An example of how a special move can be implemented is seen in the snippet below where promotion is modelled. First, the local definitions \texttt{last\_y} and \texttt{promotable} \texttt{\_types} are defined. A set containing the \texttt{promotion\_squares} is constructed, which is simply the set containing the \texttt{Coordinate}s that the pawn could normally move to that are also placed on the last rank. The polymorphic helper function \texttt{sets\_combine\_} \texttt{tuple} is finally used to generate a set of tuples with combinations of \texttt{promotable\_types} and \texttt{promotion\_squares}, which can be used to return the set of promotion \texttt{Move}s.

\begin{lstlisting}
pawn_promotion: BoardState * PM`Piece -> set of Move
pawn_promotion(board_state, pawn) == (
    let last_y = if pawn.color = <white> then 8 else 1,
        promotable_types = {<knight>, <bishop>, <rook>, <queen>} in
        let promotion_squares = {coord | coord in set
          PM`type_based_moves(pawn, PM`pieces_to_obstacles(board_state))
            & coord.y = last_y} in
            {mk_Move(pawn, mk_PM`Piece(t_c_tuple.#1, t_c_tuple.#2, pawn.color)) |
              t_c_tuple in set sets_combine_tuple[PM`PieceType, PM`Coordinate]
              (promotable_types, promotion_squares)}
)
pre pawn.type = <pawn> and pawn in set board_state;
\end{lstlisting}

So far the functions have focused on how the valid moves could be determined. Performing a move is done similarly through the function \texttt{move} where it is necessary to have different behaviour for castling and en passant as the former changes the \texttt{PieceType} and the latter captures a \texttt{Piece} without moving to the square. The other types of moves can be modelled through \texttt{move\_other}. The precondition specifies that the \texttt{Move} must be valid and the postcondition specifies that the returned \texttt{Board} has a different \texttt{BoardState} and a longer \texttt{History}.

\begin{lstlisting}
public move: Board * Move -> Board
move(board, mov) == 
  if mov.from_.type = <king> and iss_castling(board, mov) 
  then move_castling(board, mov)
  elseif (mov.from_.type = <pawn> and iss_en_passant(board, mov)) 
  then move_en_passant(board, mov)
  else move_other(board, mov)
pre mov in set possible_moves(board, mov.from_)
  and mov.from_ in set board.board_state
post len board.history < len RESULT.history
  and board.board_state <> RESULT.board_state;
\end{lstlisting}

The definition of \texttt{move\_other} is seen below. First, the potentially captured piece is found which is defined in the local definition \texttt{dead\_piece}. Due to the invariant on \texttt{BoardState} it is guaranteed to contain a single \texttt{Piece} or be the empty set. The new \texttt{BoardState} can then be defined in \texttt{new\_state} as the previous state without the \texttt{dead\_piece} and with an updated version of the moved \texttt{Piece}. Finally, the new \texttt{Board} is returned.

\begin{lstlisting}
move_other: Board * Move -> Board
move_other(board, mov) == (
  let dead_piece = {p | p in set board.board_state & p.square = mov.to_.square} in
      let new_state = (board.board_state \
        (dead_piece union {mov.from_})) union {mov.to_} in
          mk_Board(new_state, [mov] ^ board.history)
)
pre pre_move(board, mov)
post post_move(board, mov, RESULT);
\end{lstlisting}

Additionally, a helper function \texttt{default\_board} can be made that defines a board with the initial position as seen in \cref{fig:coremodel:board_initial} and an empty \texttt{History}. This is used as the starting point for new games.

\begin{lstlisting}
public default_board : () -> Board
default_board() ==
(
    let board_state : BoardState = dunion {
      {mk_PM`Piece(<pawn>, mk_PM`Coordinate(x, 2), <white>) | x in set {1,...,8}},
      {mk_PM`Piece(<pawn>, mk_PM`Coordinate(x, 7), <black>) | x in set {1,...,8}}
      -- Repeat for other PieceTypes
    } in
        mk_Board(board_state, [])
);
\end{lstlisting}

\subsection{GameModule}
The GameModule (\texttt{GM}) class has the responsibility of controlling who has the current turn and declaring the game-winner. The module defines the optional union of quote types \texttt{Winner} with the values \texttt{[PM`Color | <remis>]} where \texttt{nil} indicates that the game is ongoing. Furthermore, the module defines the composite type \texttt{Game} which contains a \texttt{Board} and a \texttt{PM`Color} indicating the turn.

The function \texttt{move} is used to perform a \texttt{Move} on the \texttt{Board} and potentially determine the winner. First, the \texttt{Move} is performed through \texttt{BM`move} and saved in the local definition \texttt{new\_board}. Then it is checked whether the opponent has any valid \texttt{Move}s for the next turn. If not then the \texttt{Game} is either won by the player or ended in remis, depending on whether the opponent is in check.

\begin{lstlisting}
public move : Game * BM`Move -> (Game * Winner)
move(game, mov) == (
  let new_board = BM`move(game.board, mov),
    opposite_c = PM`opposite_color(game.turn) in
    if forall p in set new_board.board_state &
      p.color = opposite_c => BM`possible_moves(new_board, p) = {} then
      if BM`in_check(new_board.board_state, opposite_c) then
        mk_(game, game.turn)
      else
        mk_(game, <remis>)
    else
      mk_(mk_Game(new_board, opposite_c), nil)
)
pre mov.from.color = game.turn and
  mov in set BM`possible_moves(game.board, mov.from)
post len game.board.history < len RESULT.#1.board.history
    and game.board.board_state <> RESULT.#1.board.board_state;
\end{lstlisting}

\subsection{Runner}
The last part of the model consists of a class, \texttt{Runner}, that reads the contents of a PGN file (\cref{sec:PGN}), converts it to \texttt{Move}s through the PGN module, and iteratively performs the moves.

From a \texttt{Runner} perspective it could also potentially be interesting to create a graphical rendering of the Chess board itself. This could potentially be achieved using either VDMPad \cite{Oda&15a} or ViennaTalk \cite{Oda&16a}. In an Overture context this kind of visualisation was also enabled when it was based on Eclipse but this has not yet been fully incorporated in the VSC version \cite{Nielsen&12}.

\section{Portable Game Notation}
\label{sec:PGN}


The Portable Game Notation~\cite{PortableGameNotation2022} (PGN) is the de-facto standard used for chess annotation on many online chess websites.\footnote{PGN is supported on websites such as \url{chess.com} \url{https://lichess.org/}, and \url{chess24.com}.}
PGN consists of information related to the chess game (e.g. player information) and move text, where the move text describes the actual piece moves of the game. PGN uses letters for the x-axis and numbers for the y-axis as described in \cref{sec:background}. Generally speaking, a move in PGN consists of the \texttt{PieceType} as the first character and the \texttt{Coordinate} as the second and third characters. Furthermore, there is special notation for castling, check, checkmate, and extra information may be added to remove ambiguity.

PGN was added as a way to verify the overall integrity of the VDM model by testing it on some real-world data. Since VDM++ does not include a string manipulation library it was necessary to partly define one. Furthermore, the PGN class has the responsibility of parsing a valid \texttt{String} describing a game of chess through PGN notation to the VDM++ \texttt{Move} representation and vice versa.

\begin{lstlisting}
values
 numerical_chars = "0123456789";
 numerical_char_to_nat : inmap char to nat = 
             {numerical_chars(i) |-> i-1 | i in seq numerical_chars};

 valid_x_chars = "abcdefgh";
 x_char_to_nat1 : inmap char to nat1 = 
              {valid_x_chars(i) |-> i | i in seq valid_x_chars};

 piece_type_to_string : inmap PM`PieceType to String = 
                        {<pawn> |-> "", <rook> |-> "R", <knight> |-> "N",
                         <bishop> |-> "B", <queen> |-> "Q", <king> |-> "K"}

functions

public move_to_pgn_string: BM`Move -> String
move_to_pgn_string(move) ==
    let piece_type = piece_type_to_string(move.from.type),
        x = (inverse x_char_to_nat1)(move.to_.square.x),
        y = (inverse numerical_char_to_nat)(move.to_.square.y) 
     in
        piece_type ^ [x] ^ [y];
\end{lstlisting}

The helper function \texttt{move\_to\_pgn\_string} is used in \texttt{Runner} when parsing a chess game and logging the results to a text file. The \texttt{PieceType} is mapped to a string by applying the type to \texttt{piece\_type\_to\_string}. Similarly, the x- and y-coordinates are found but the inverse mappings are used here, which is possible as the maps are injective. 

This class is currently made in VDM but since VDM is not really meant for parsing strings it would make sense to redo this part in Java as a new library for reading this external format directly into VDM structures. It could even use scanner and parser generators inside if desired but this is mainly an issue about the potential error reporting in case the string provided does not live up to the PGN syntax.

\section{Related Work}
\label{sec:related}


The rules of the chess game have been incorporated in other formalisms as well but from the different publications it is unclear to what extend they have incorporated the more complicated rules such as castling \cite{Korner&21,Krings&21,Saralaya&11}. Most of the other formal methods analysis of the chess game are considering this from a model checking perspective. Thus, the focus is primarily in analysing the potential future outcome of a game of chess or simply looking into whether a sequence of moves are legal or not. In contrast the work presented in this paper focus more on the most natural way to represent the rules of the chess game. However, in \cite{Krings&21} there is an attempt of first generalising the rules of games before specialising it to different games such as chess. It is possible that this approach could be reused more systematically in the work presented in this paper with advantage as well. 

\section{Concluding Remarks and Future Work}
\label{sec:concl}

This article presented a feature-complete model of Chess in VDM++ that can be used as an educational example or as a basis for formal analysis of other topics related to the game of Chess. The model has been added as one of the models that can be imported into the Visual Studio Code version of Overture \cite{Lund&22} so others can also experiment with the entire VDM++ model. 

In general, we find that there are interesting pros and cons when considering which paradigm to use for modelling a system in VDM++. In the case of the game of Chess, we find that the functional paradigm works better than the object-oriented paradigm (OOP), but it also has drawbacks. If OOP is chosen for a given system, one must be aware of the complications it may bring, in particular relating to invariants across object references. The same challenge would also appear for operations with post-conditions because this also could relate to instance variables in other objects before execution of the operation.

\paragraph{Acknowledgements}
We would like to thank the anonymous reviewers for valuable feedback on the original version of this paper.

\bibliographystyle{splncs04}
\bibliography{references.bib}

\end{document}